\documentclass[conference,a4paper]{IEEEtran}
\IEEEoverridecommandlockouts
\usepackage{cite}
\usepackage{amsmath,amssymb,amsfonts}
\usepackage{algorithmic}
\usepackage{graphicx}
\usepackage{textcomp}
\usepackage{xcolor}
\def\BibTeX{{\rm B\kern-.05em{\sc i\kern-.025em b}\kern-.08emT\kern-.1667em\lower.7ex\hbox{E}\kern-.125emX}}
\begin{document}

\title{Correlation Analysis between Terrain Features and eLoran Spatial ASF according to DEM Spatial Resolution
\thanks{This work was supported in part by Grant RS-2024-00407003 from the ``Development of Advanced Technology for Terrestrial Radionavigation System'' project, funded by the Ministry of Oceans and Fisheries, Republic of Korea;
in part by the National Research Foundation of Korea (NRF), funded by the Korean government (Ministry of Science and ICT), under Grant RS-2024-00358298; 
in part by the Future Space Navigation and Satellite Research Center through the NRF, funded by the Ministry of Science and ICT (MSIT), Republic of Korea, under Grant 2022M1A3C2074404; 
and in part by the MSIT, Korea, under the Information Technology Research Center (ITRC) support program supervised by the Institute of Information \& Communications Technology Planning \& Evaluation (IITP) under Grant IITP-2024-RS-2024-00437494.
}
}

\author{\IEEEauthorblockN{Jaewon Yu} 
\IEEEauthorblockA{\textit{School of Integrated Technology} \\
\textit{Yonsei University}\\
Incheon, Korea \\
jaewon.yu@yonsei.ac.kr} 
\and
\IEEEauthorblockN{Pyo-Woong Son${}^{*}$} 
\IEEEauthorblockA{\textit{Department of Electronics Engineering} \\
\textit{Chungbuk National University} \\
Cheongju, Korea \\
pwson@cbnu.ac.kr}
{\small${}^{*}$ Corresponding author}
}

\maketitle

\begin{abstract}
The ASF is a major source of positioning error in eLoran system and is affected by terrain characteristics along the signal propagation path.
This study analyzes the correlations between spatial ASF and path-based terrain features using NASA SRTM 30m and NGII 90m DEMs.
Path mean elevation and path mean slope were extracted along the propagation paths from the Gwangju eLoran transmitter to measurement locations in the Incheon and Pyeongtaek port areas.
The results show that both terrain features are strongly correlated with spatial ASF, with the higher-resolution NASA SRTM DEM generally exhibiting stronger correlations, particularly for path mean slope.
These findings demonstrate the importance of DEM spatial resolution in terrain-based spatial ASF analysis and provide a basis for selecting terrain data for future machine-learning-based ASF prediction.

\end{abstract}


\section{Introduction}
Global Navigation Satellite System (GNSS) \cite{Kim14:Comprehensive, Chen11:Real, Lee23:Seamless, Kim23:Machine, Kim23:Low, Lee24:A, Kim23:Single, Kim22:Machine, Lee22:Urban, Kim25:Set, Jeong24:Quantum} have become a fundamental infrastructure for modern positioning, navigation, and timing (PNT) services.
However, because GNSS signals are received at low power levels, they are inherently vulnerable to intentional radio-frequency interference, such as jamming and spoofing \cite{Park21:Single, Park18:Dual, Kim19:Mitigation, Park17:Adaptive, Jeong20:RSS, Moon24:HELPS, Lee22:Performance, Park25:Toward}.
This vulnerability has increased the need for robust and independent backup PNT systems, and enhanced Loran (eLoran) \cite{Yin25:Preliminary,Yang25:Research,Cheng25:Research}, a high-power low-frequency terrestrial navigation system, has received considerable attention as a resilient alternative to GNSS \cite{Son20:eLoran}.

In particular, GPS interference originating from North Korea has occurred repeatedly on the Korean Peninsula, with both its frequency and impact increasing in recent years.
The number of reported GPS interference incidents increased from 39 in 2023 to 578 in 2024, representing an approximately fifteen-fold increase \cite{Kelleher25:Jamming}.
Furthermore, large-scale GPS interference between May 29 and June 2, 2025 affected over 500 aircraft and hundreds of vessels \cite{Kelleher25:Jamming}.
In response to repeated GNSS interference, the International Civil Aviation Organization (ICAO) adopted Resolution A42-3 in October 2025 concerning recurring GNSS radio-frequency interference originating from North Korea \cite{ICAO25:A42}.
To enhance resilience against such threats, Korea has been developing eLoran infrastructure as a terrestrial backup navigation system \cite{Son22:Compensation,Son23:Demonstration}.
Similarly, the United Kingdom announced a \pounds155 million investment in national PNT resilience in November 2025, including \pounds71 million for a national eLoran program \cite{DSIT25:PNT}.

The propagation delay of an eLoran signal from a transmitter to a receiver can be divided into the Primary Factor (PF), Secondary Factor (SF), and Additional Secondary Factor (ASF).
PF represents the propagation delay associated with the propagation velocity of the signal in the atmosphere, whereas SF represents the additional delay that occurs when the signal propagates over a uniform sea surface.
The propagation delay associated with PF and SF can be calculated with relatively high accuracy as a function of propagation distance using the Brunavs model.
In contrast, ASF represents the additional propagation delay caused by variations in terrain and surface properties along the actual propagation path \cite{Hehenkamp2023:Prediction}.
Because these properties vary spatially, ASF exhibits complex spatial variations and is difficult to predict accurately using simple theoretical models.
Consequently, ASF is one of the major error sources limiting eLoran positioning accuracy \cite{Rhee21:Enhanced}.

ASF variations can be further characterized in terms of temporal ASF and spatial ASF.
Temporal ASF represents time-varying changes in ASF caused by environmental factors such as weather conditions and can be compensated for in real time using ASF measurements from nearby reference stations.
In contrast, spatial ASF varies with location but remains relatively stable over time at a given location, allowing it to be compensated for using a pre-generated ASF map \cite{Kim22:First}.

Various approaches have therefore been investigated to generate spatial ASF maps from measurements collected at discrete locations \cite{Williams00:ASF,Hwang18:TDOA,Son19:Universal}.
In particular, Universal Kriging has been applied to generate spatially continuous ASF maps \cite{Son19:Universal}, while Ordinary Kriging has also been investigated for regional ASF prediction \cite{Di24:Regional}.
Field measurements of long-wave propagation have demonstrated significant temporal and spatial correlations in propagation delays between nearby locations, while meteorological conditions, ground electrical properties, and terrain variations along the propagation path have been identified as relevant influencing factors \cite{Di25:Longwave}.
Environmental and ground-surface information has also been investigated for predicting ground-wave propagation delays in terrestrial radionavigation systems \cite{Hehenkamp2023:Prediction}.

More recently, data-driven approaches have been introduced for ASF analysis and prediction.
Machine-learning-based methods using meteorological and terrain information as input features have been proposed to improve eLoran timing accuracy \cite{Kang25:Enhancing}.
In such approaches, the characteristics and quality of the input data can directly affect model training and prediction performance.
However, the effect of the spatial resolution of terrain data on the relationship between terrain characteristics and spatial ASF has not yet been sufficiently investigated.

In this study, we quantitatively analyze the relationship between propagation-path terrain characteristics and spatial ASF using two digital elevation models (DEMs) with different spatial resolutions.
Measured spatial ASF maps from the Incheon and Pyeongtaek port areas are merged to construct the reference dataset, and a propagation path from the Gwangju eLoran transmitter to each ASF measurement location is generated.
For each propagation path, the mean elevation and mean terrain slope are extracted and their correlations with spatial ASF are evaluated.
The same analysis is performed using DEM data at two different spatial resolutions, 30m and 90m, to evaluate the effect of DEM spatial resolution on the resulting correlations.
Through this analysis, we identify terrain characteristics that exhibit significant correlations with spatial ASF and evaluate how DEM spatial resolution affects the observed relationships between terrain characteristics and spatial ASF.
The results provide a basis for selecting terrain input data and DEM spatial resolution for future machine-learning-based spatial ASF prediction models.

\section{Methodology}

\subsection{Spatial ASF Map and Multi-Resolution DEM Dataset}

\begin{figure}
    \centering
    \includegraphics[width=0.8\linewidth]{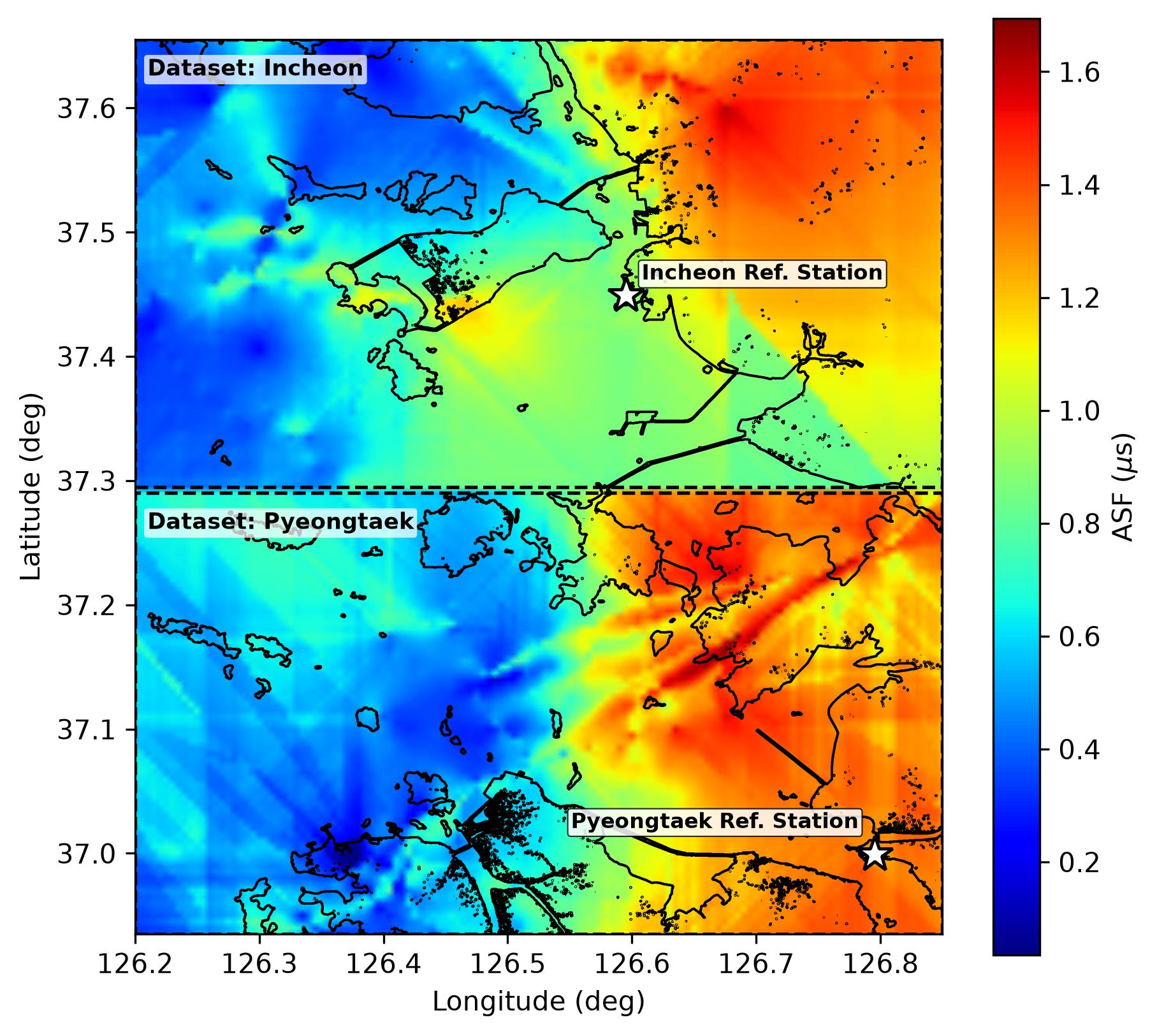}
    \caption{Measured Spatial ASF Map of the merged Incheon and Pyeongtaek dataset (Gwangju/9930W transmitter).}
    \label{fig:asfmap}
\end{figure}

In this study, measured Spatial ASF maps \cite{NMPO23:ASF} of the Incheon and Pyeongtaek regions were used as ground truth.
Fig.~\ref{fig:asfmap} shows the merged spatial ASF map of Incheon and Pyeongtaek used in this study.
Spatial ASF maps were provided by the Korea Research Institute of Ships and Ocean Engineering(KRISO) in the RTCM SC-127 format, generated with respect to the Gwangju (9930W) eLoran transmitter.
The grid consists of 145 points in latitude and 131 points in longitude, for a total of
18,995 points in the grid, spanning a latitude range of $36.935$--$37.655^{\circ}$N and a longitude
range of $126.200$--$126.850^{\circ}$E, with a grid spacing of $0.005^{\circ}$ (approximately 500m).
The ASF values are given in $\mu$s.

The terrain data consist of a 30m-resolution NASA Shuttle Radar Topography Mission (SRTM) DEM \cite{NASA:SRTM} and a 90m DEM provided by the
National Geographic Information Institute (NGII) of Korea \cite{NGII:DEM}.
The NASA SRTM DEM is provided in a Geographic Coordinate System (GeoTIFF) format, whereas the NGII DEM
is provided in a Projected Coordinate System based on EPSG:5186.
Both DEMs were used at their native resolution and the same analysis procedure was applied to each.

\subsection{Terrain Feature Extraction along the Propagation Path}

Spatial ASF is affected not only by the terrain at the receiver location but also by the terrain along the entire propagation path between the transmitter and receiver. 
Therefore, for each ASF measurement point, a great-circle propagation path was generated between the Gwangju eLoran transmitter and the receiver, and elevation and terrain slope data were extracted from the DEM along the corresponding path.

The NASA SRTM DEM is provided as a GeoTIFF in a geographic coordinate system (EPSG:4326), with elevation values defined on a regular grid at an approximately 30 m spatial resolution. 
In contrast, the NGII DEM is provided in a projected coordinate system (EPSG:5186), with elevation values defined on a regular grid at a 90 m spatial resolution. 
In both DEMs, elevation values are available only at discrete grid locations. 
Therefore, the ASF measurement points and the sample points generated along the propagation paths do not necessarily coincide with the DEM grid points. 
To estimate elevation values at these locations, bilinear interpolation was applied using the four surrounding DEM grid points.

Using the interpolated terrain data, the following path-based terrain features were calculated: (1) path mean elevation, defined as the average of the elevation values sampled along the propagation path; and (2) path mean slope, defined as the average of the terrain slope values sampled along the propagation path. 
Terrain slope was calculated from the numerical gradient of each DEM at its native spatial resolution and expressed in degrees.

\section{Results}

\subsection{Path Mean Elevation and Correlation with Spatial ASF}

\begin{figure}[t]
\centering
\includegraphics[width=0.9\columnwidth]{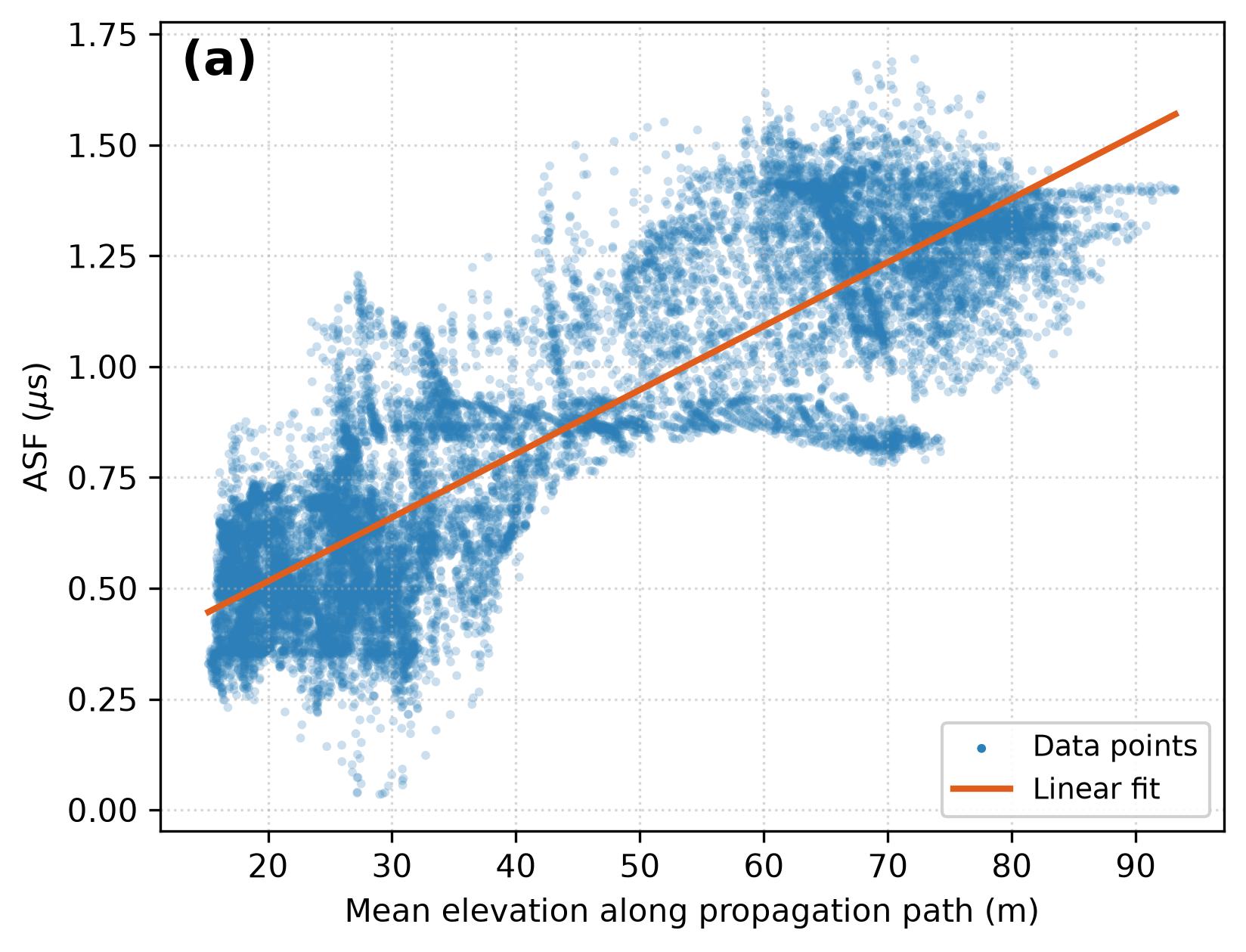}
\vspace{2pt}
\includegraphics[width=0.9\columnwidth]{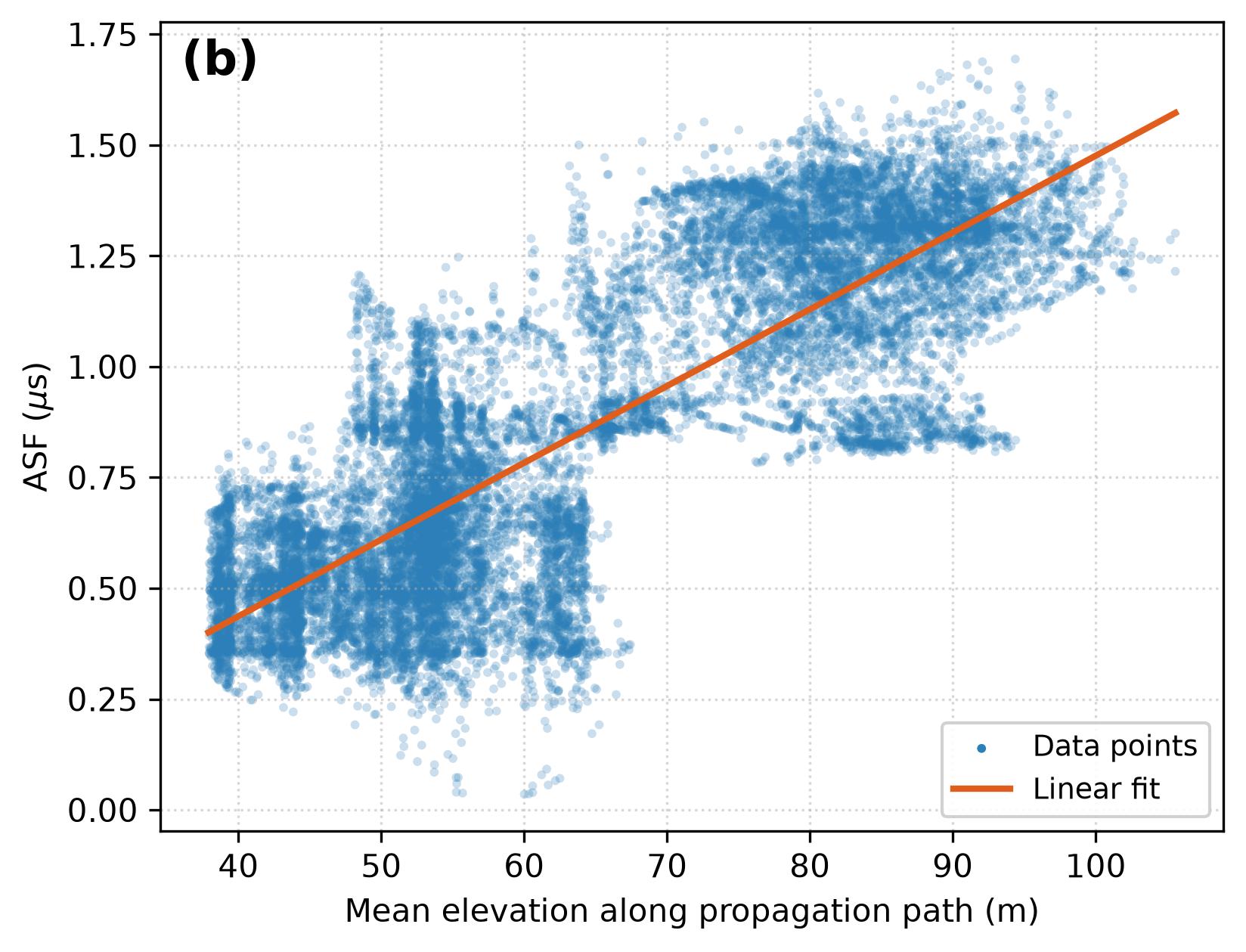}
\caption{Path mean elevation vs. Spatial ASF using (a) NASA SRTM 30m DEM and (b) NGII 90m DEM (merged dataset).}
\label{fig:fig4}
\end{figure}

Fig.~\ref{fig:fig4} shows the scatter plots between the path mean elevation and spatial ASF, and
Table~\ref{tab:pathmean} summarizes the Pearson and Spearman correlation coefficients for each DEM.

\begin{table}
\centering
\caption{Correlation between Path Mean Elevation and Spatial ASF}
\begin{tabular}{|l|c|c|c|c|}
\hline
\textbf{DEM} & \textbf{Pearson $r$} & \textbf{$p$} & \textbf{Spearman $r$} & \textbf{$p$} \\
\hline
NASA SRTM 30m & 0.870 & $<0.001$ & 0.821 & $<0.001$ \\
NGII 90m & 0.827 & $<0.001$ & 0.780 & $<0.001$ \\
\hline
\end{tabular}
\label{tab:pathmean}
\end{table}

A strong correlation was confirmed between the path mean elevation and spatial ASF.
The Pearson correlation coefficient for the NASA SRTM DEM was $0.870$, and the Spearman correlation coefficient was $0.821$, while for the NGII DEM the corresponding values were $0.827$ and $0.780$, respectively.
This confirms that the terrain information along the entire propagation path is effective in explaining spatial ASF, and the path mean elevation computed using the higher-resolution NASA SRTM DEM showed a higher correlation with spatial ASF than that computed using the NGII DEM.

\subsection{Path Mean Slope and Correlation with Spatial ASF}
\label{sec:slope}

\begin{figure}[t]
\centering
\includegraphics[width=0.9\columnwidth]{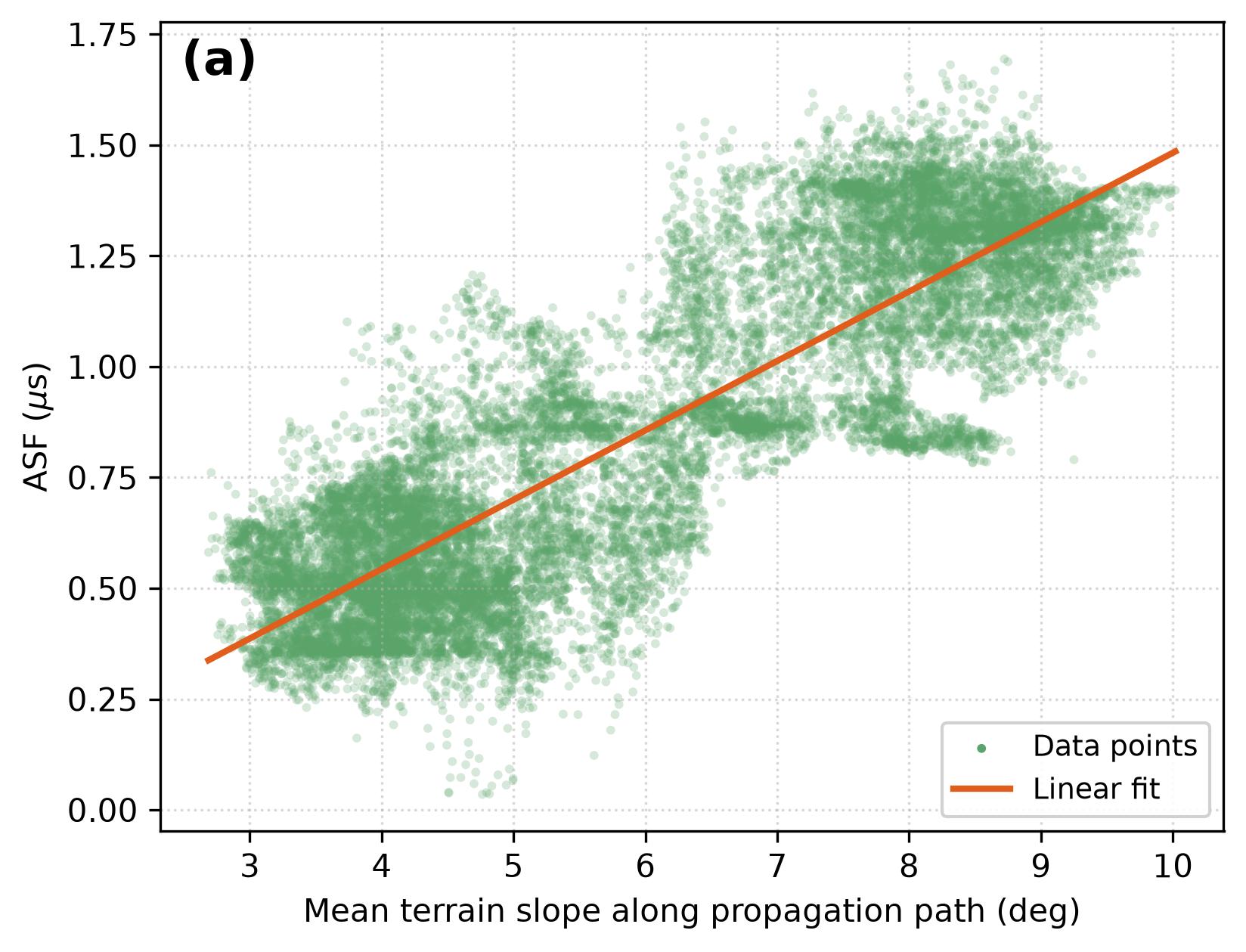}
\vspace{2pt}
\includegraphics[width=0.9\columnwidth]{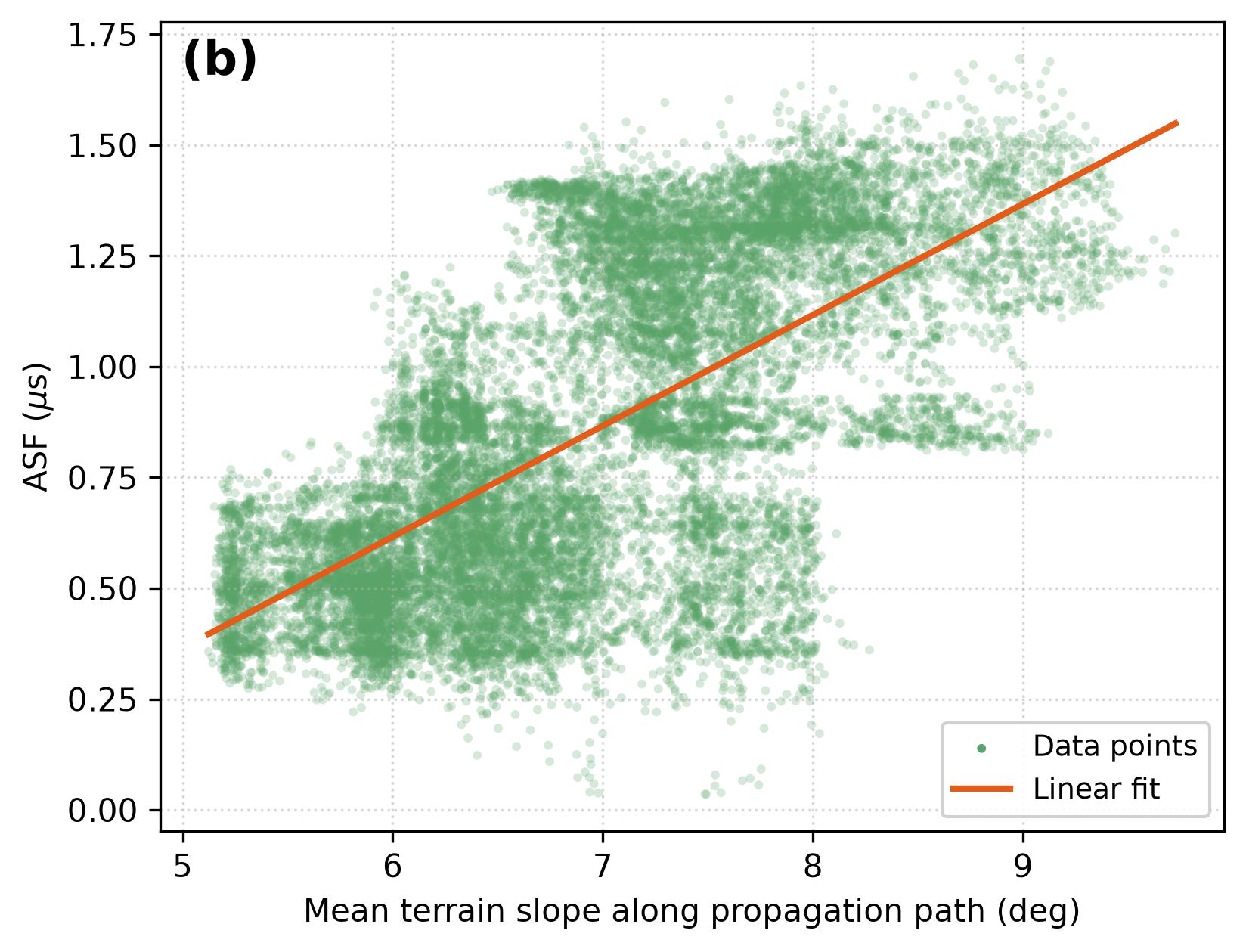}
\caption{Path mean slope vs. Spatial ASF using (a) NASA SRTM 30m DEM and (b) NGII 90m DEM (merged dataset).}
\label{fig:fig6}
\end{figure}

In addition to the elevation values themselves, terrain slope, which indicates how abruptly
elevation changes, is another terrain feature that can influence spatial ASF.
In this study, we analyzed the correlation using the path mean slope, which
reflects the entire propagation path.
Fig.~\ref{fig:fig6} shows the scatter plots between the path mean slope and spatial ASF, and Table~\ref{tab:meanslope} summarizes the Pearson and Spearman correlation coefficients for each DEM.

\begin{table}
\centering
\caption{Correlation between Path Mean Slope and Spatial ASF}
\begin{tabular}{|l|c|c|c|c|}
\hline
\textbf{DEM} & \textbf{Pearson $r$} & \textbf{$p$} & \textbf{Spearman $r$} & \textbf{$p$} \\
\hline
NASA SRTM 30m & 0.858 & $<0.001$ & 0.814 & $<0.001$ \\
NGII 90m & 0.651 & $<0.001$ & 0.641 & $<0.001$ \\
\hline
\end{tabular}
\label{tab:meanslope}
\end{table}

As shown in Table~\ref{tab:meanslope}, the path mean slope showed a strong correlation with spatial ASF for both DEMs.
The Pearson correlation coefficient for the NASA SRTM DEM was $0.858$, and the Spearman correlation coefficient was $0.814$, while for the NGII DEM the corresponding values were $0.651$ and $0.641$, respectively.
As with the path mean elevation, the path mean slope computed using the higher-resolution NASA SRTM DEM showed a higher correlation with spatial ASF than that computed using the NGII DEM.
The Pearson correlation coefficient difference between NASA SRTM and NGII was $0.858-0.651=0.207$ for the path mean slope, more than four times larger than the corresponding difference of $0.870-0.827=0.043$ for the path mean elevation.

This is interpreted as resulting from the fact that slope is a derivative feature representing the spatial rate of change of elevation, so DEMs with coarser grid spacing smooth out fine-scale terrain undulations, substantially reducing their ability to reflect actual terrain variation.
In contrast, the path mean elevation reflects the absolute elevation level over a broad section of the path rather than local undulations, and therefore appears to be relatively less sensitive to the degradation of DEM resolution.

\section{Conclusion}

In this study, the correlations between spatial ASF and path-based terrain features were analyzed using measured spatial ASF maps and DEM data from Incheon and Pyeongtaek.

Path mean elevation and path mean slope showed strong and statistically significant correlations with spatial ASF. 
The higher-resolution NASA SRTM 30m DEM generally showed stronger correlations than the NGII 90 m DEM, particularly for path mean slope. 
This indicates that DEM spatial resolution has a greater influence on slope-related features, which are more sensitive to local terrain variations, and highlights the importance of high-resolution terrain data for machine-learning-based spatial ASF prediction.

Future work will incorporate effective ground conductivity derived from land-cover data with elevation and slope features to develop a machine-learning model for Spatial ASF map prediction.

\section*{Acknowledgment}

Generative AI (ChatGPT, OpenAI) was used solely to assist with grammar and language improvements during the manuscript preparation process.  
No content, ideas, data, or citations were generated by AI.  
All technical content, methodology, analysis, and conclusions were written and verified solely by the authors.

\bibliographystyle{IEEEtran}
\bibliography{mybibfile, IUS_publications}

\vspace{12pt}

\end{document}